\documentclass[journal]{IEEEtran}
\usepackage{graphicx}

\usepackage{cite}      

\usepackage{subfigure} 

\begin{document}

%
\title{A Simple Distributed Antenna Processing Scheme for Cooperative Diversity}

\author{Yijia~Fan, Abdulkareem~Adinoyi, John~S~Thompson,
Halim~Yanikomeroglu, H.~Vincent~Poor
\thanks{Part of this work has appeared in European Wireless Conference 2006, Athens, Greece
 and International Conference on Communications, Glasgow, UK,
 2007.}
\thanks{Y. Fan and H. V. Poor are with Department of Electrical
Engineering, Princeton University, Princeton, NJ 08544
(email:\{yijiafan,poor\}@princeton.edu).} \thanks{J. S. Thompson is
with the Institute for Digital Communications, University of
Edinburgh, Edinburgh, EH9 3JL, UK (e-mail: john.thompson@ed.ac.uk).}
\thanks{A. Adinoyi and H. Yanikomeroglu are with Broadband
Communications and Wireless Systems (BCWS) Centre, Department of
Systems and Computer Engineering, Carleton University, Ottawa, K1S
5B6, Canada (Email: \{halim.yanikomeroglu,
adinoyi\}@sce.carleton.ca).}
\thanks{This research was supported in
part by the U. S. National Science Foundation under Grants
ANI-03-38807 and CNS-06-25637.}}

\maketitle

\begin{abstract}
In this \emph{letter} the performance of multiple relay channels is
analyzed for the situation in which multiple antennas are deployed
only at the relays. The simple repetition-coded decode-and-forward
protocol with two different antenna processing techniques at the
relays is investigated. The antenna combining techniques are maximum
ratio combining (MRC) for reception and transmit beamforming (TB)
for transmission. It is shown that these distributed antenna
combining techniques can exploit the full spatial diversity of the
relay channels \emph{regardless of} the number of relays and
antennas at each relay, and offer \emph{significant} power gain over
distributed space-time coding techniques.
\end{abstract}


%

\section{Introduction}
%
%
%
%

The performance limits of distributed space-time codes, which can
exploit \emph{cooperative diversity}, have been investigated in
\cite{23} and \cite{7} for single-antenna relay networks using
random coding techniques. However, the design and implementation of
practical codes that approach these limits are challenging open
research areas. One approach to these problems might be to use known
space-time codes for the point-to-point multiple-input
multiple-output (MIMO) link (e.g. \cite{stdo}) in relay networks.
However, the processing complexity at each relay node for such an
approach can increase significantly, as antennas in relay networks
are distributed rather than centralized. For example, each relay may
need to know all of the uncoded data, before sending only one part
of the codeword to the destination. Similarly, the decoding process
at the destination might also be very complex when the number of
relays are large. Moreover, a more complex protocol is required in
order to assign different relays to transmit different parts of the
codeword. These points lead to additional time delay and energy
cost, while they also present fundamental issues especially for
large ad-hoc or sensor networks \cite{dana,18,vince}. Simpler codes
such as space-time block codes \cite{stbd} will result in a rate
loss when the number of relays is more than two. Another recent
scheme exploits the selection diversity of the network by selecting
the best relay among all the available relays \cite{sel}. However,
the power gain for this scheme is limited due to the limited power
at a single relay node; especially in a sensor network environment.

In this \emph{letter} we exploit the spatial diversity of relay
channels in an alternative way to the space-time codes-based
approach. We apply two kinds of antenna processing techniques at the
relay, namely maximum ratio combining (MRC) \cite{mrc} for reception
and transmit beamforming (TB)\cite{29} for transmission. These
techniques are often used in point-to-point single-input
multiple-output (SIMO) or multiple-input single-output (MISO)
wireless links and have been shown to achieve the optimal diversity
multiplexing tradeoff in these cases\cite{fund}. More specifically,
for a MISO channel, beamforming is often considered as a better
approach than space-time coding due to its higher \emph{power}
\emph{gain}, provided that the channel state information (CSI) can
be fed back to the transmitter. In our model, we move the multiple
antennas to the relays, while the source and the destination are
equipped with only a single antenna. \emph{Unlike} the
point-to-point link, the antennas are deployed in a distributed
fashion, and MRC and beamforming can only be performed in a
distributed rather than a centralized fashion in this scenario. One
of the contributions of this letter is to investigate the diversity
and power performance tradeoff between the number of relays and the
number of antennas at each relay. We will also compare distributed
MRC-TB with space-time coding in a multi-antenna multi-relay
environment. Some related work on single antenna relay networks has
also considered beamforming approach, although this earlier work
focuses primarily on the energy efficiency or capacity scaling
behavior of such networks \cite{dana,bwang}.


\section{System description}

\begin{figure}[t!]
\centering
\includegraphics[width=2.5in]{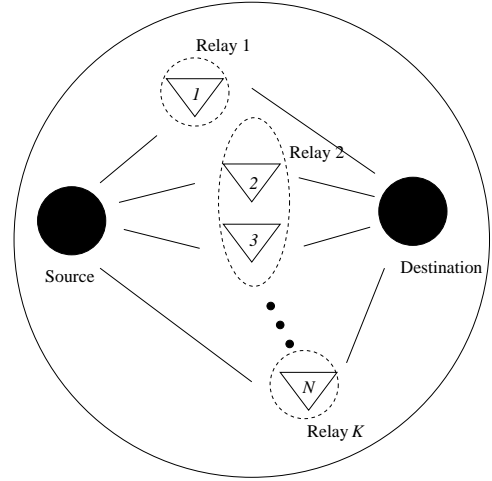}
\caption{System model for a two hop network: The source and
destination are each deployed with one antenna. Totally $N$ antennas
are deployed at $K$ relays. For each channel realization, all the
backward or forward channel coefficients for all $N$ antennas remain
the same regardless of the number of relays $K$. }
\label{coll_model}
\end{figure}

We consider a two hop network model with one source, one destination
and $K$ relays. For simplicity we ignore the direct link between the
source and the destination. The extension of our results to include
the direct link is straightforward. We assume that the source and
destination are deployed with a single antenna, while relay $k$ is
deployed with $m _k$ antennas; the total number of antennas at all
relays is fixed to $N$. We restrict attention to the case in which
the channels exhibit slow and frequency-flat fading. We assume a
coherent relay channel configuration context in which the $k$th
relay can obtain full knowledge of both the backward channel vector
${\bf{h}}_k$ and the forward channel vector ${\bf{g}}_k$. Note that
forward channel knowledge can be obtained easily if the
relay-destination link operates in a Time-Division-Duplex (TDD)
mode. One example where the relays obtain the required channel
information can be found in \cite{bwang}, but this might require
additional signalling overhead. In a slow fading channel, which is
the focus of this \emph{letter}, this overhead is negligible. For
fair comparison, we also assume that for each channel realization,
all the backward and forward channel coefficients for all $N$
antennas remain the same regardless of the number of relays $K$.
Fig. \ref{coll_model} shows the system model.

Data is transmitted over two time slots using two hops. In the first
transmission time slot, the source broadcasts its signal to all
relay terminals. The input/output relation for the source to the
$k$th relay is given by \setlength{\arraycolsep}{0.0em}
\begin{equation}
{\bf{r}}_k  = \sqrt{\eta} {\bf{h}}_k s + {\bf{n}}_k, \label{1}
\end{equation}
where ${\bf{r}}_k$ is the $m _k \times 1$ received signal vector,
$\eta$ is the transmit power at the source, $s$ is the unit mean
power transmitted signal, and ${\bf{n}}_k$ is $m_ k \times1$ complex
circular additive white Gaussian noise at relay $k$ with zero mean
and identity covariance matrix ${\bf{I}}_{m_k}$. The entries of the
channel vector ${\bf{h}}_k$ are independent and identically
distributed (i.i.d.) complex Gaussian random variables with zero
means and unit variances. We assume that each relay performs MRC of
the received signals, by multiplying the received signal vector by
the vector ${{{{\bf{h}}_k^H } \mathord{\left/
 {\vphantom {{{\bf{h}}_k^H } {\left\| {{\bf{h}}_k } \right\|_F }}} \right.
 \kern-\nulldelimiterspace} {\left\| {{\bf{h}}_k } \right\|_F }}}$,
 where $\left\| \bullet \right\|_F$ denotes the Frobenius norm.
 The SNR at the output of the receiver in this scenario can be
written as
\begin{equation}
\rho _k^{\left(m _k\right)}  = \eta \sum\limits_{i = 1}^{m_k }
{\left| {h_{i,k} } \right|^2 }, \label{eqn:SNRmrc}
\end{equation}
where $h_{i,k}$ denotes the channel coefficient from the source to
the $i$th antenna at relay $k$. Note that for space-time coding, the
same MRC scheme is used at the relays when comparing with
distributed MRC-TB later in this \emph{letter}.

After the relays decode the signals, each relay re-encodes the
signal using the same codebook as used at the source, then performs
TB of the decoded waveform. If we denote the unit variance
re-encoded signals as $t _k$, the transmitted signal vector
${\bf{d}}_k$ for relay $k$ can be written as
\begin{equation}
{\bf{d}}_k  = \sqrt {\frac{{\eta m_k }}{N}} \frac{{{\bf{g}}_k^H
}}{{\left\| {{\bf{g}}_k } \right\|_F }}t_k,
 \label{eqn:tb}
\end{equation}
where the vector ${{\bf{g}}_k }$ is the $m_k \times 1$ channel
vector from the $k$th relay to the destination, where components are
i.i.d. complex Gaussian random variable with zero means and unit
variances. The vector ${{\bf{d}}_k }$ in (\ref{eqn:tb}) is designed
to meet the total transmit power constraint:
\begin{equation}
{\rm E}\left[ {\left\| {{\bf{d}}_k } \right\|_F^2 } \right] \le
\frac{{\eta m_ k}}{N}. \label{3}
\end{equation}
Here we assume that the total transmit power from all relays is
fixed to be $\eta$, i.e., the same as the source transmit power.
However, all the conclusions in the paper also hold when the total
power from all relays is fixed to \emph{an arbitrary constant}. We
note that this power assumption has a meaningful practical
implication: in reality a transmitter having a larger number of
antennas can often transmit with a higher power (in proportional to
the number of transmit antennas in this paper). The destination
receiver simply decodes the combined signals from all $K$ relays. If
the signals are correctly decoded at all the relays (i.e., $t _k =
s$ for all $k$), the output SNR at the destination receiver can be
written as
\begin{equation}
\rho _d^ {\left\{m _k\right\}}  = \left( {\sum\limits_{k = 1}^K
{\sqrt {\frac{{\eta m_k }}{N}\sum\limits_{i = 1}^{m_k } {\left|
{g_{i,k} } \right|^2 } } } } \right)^2. \label{eqn:SNRtb}
\end{equation}
When each of the relays is deployed with a single antenna, there is
no MRC gain at the relays, nor is there any beamforming gain at the
destination. However, the destination still observes a set of
equal-gain-combined \cite{EGC} amplitude signals from all relays.
Since we assume that the backward and forward channel coefficients
for each antenna are kept the same for different values of $K$ and
$m _i$, the output SNR at the destination can be rewritten as $\rho
_d^{\left(1\right)}  = \frac{\eta }{N}\left( {\sum\limits_{k = 1}^K
{\sum\limits_{i = 1}^{m_i } {\left| {g_{i,k} } \right|} } }
\right)^2;$ when all the antennas are deployed on one relay (i.e.,
$K=1$ and $m _1=N$), full diversity gain is achieved among all the
$N$ antennas at the relay and also at the destination. The SNR for
this case can be rewritten as $\rho _d^{\left(N\right)} = \eta
\sum\limits_{k = 1}^K {\sum\limits_{i = 1}^{m_i } {\left| {g_{i,k} }
\right|^2 } }$.

\section{Performance Analysis}

\subsection{SNR Gain}

We first compare $\rho_d^{\left(1\right)}$ with the output SNR at
the destination when space-time coding \cite{23} is used, which can
be written as
\begin{equation}
\rho_{std}=\frac{\eta }{N}\sum\limits_{k = 1}^K {\sum\limits_{i =
1}^{m_i } {\left| {g_{i,k} } \right|^2 } }  = \frac{{\rho
_d^{\left(N\right)} }}{N}. \label{stdcomp}
\end{equation}
Clearly we can see that $\rho_d^{\left(1\right)} \ge \rho_{std}$. We
now introduce the bounds on the value of $\rho
_d^{\left\{m_k\right\} }$, for $m_k = 1\dots N$.
\newtheorem{lemma}{Lemma}
\begin{lemma}
For any $\left\{m _k\right\}$, $\rho _d^{\left(N\right)}  \ge \rho
_d^{\left\{m_k\right\} } \ge \rho _d^{\left(1\right)}$.
\end{lemma}
The proof is omitted due to space limitations; please refer to
\cite{ett} for details. This lemma implies that, generally, the
increased ``equal gain combining'' gain at the destination cannot
compensate for the loss of MRC gain at the relay and TB gain at the
destination when $K$ is increased and each $m_k$ is reduced, given
the power constraint (\ref{3}). Because $\rho_d^{\left(1\right)} \ge
\rho_{std}$, we can thus conclude that MRC-TB leads to a higher
instantaneous receive SNR than space-time coding at the destination,
again given the power constraint (\ref{3}) and the assumption that
all relays can decode the source message correctly.

\subsection{Outage Analysis}

To examine the outage properties, we begin with the following
result.

\newtheorem{theorem}{Theorem}
\begin{lemma}
Assuming that all the relays correctly decode the message, the
outage probability $P_{out}^{\left\{m_k\right\} }$ for the relay
network is approximately bounded by
\begin{equation}
\frac{1}{{N!}} \left( {\frac{{N\left( {2^{2R}  - 1} \right)}}{\eta
}} \right)^N \ge P_{out}^{\left\{m_k\right\} }  \ge \frac{1}{{N!}}
\left( {\frac{{2^{2R}  - 1}}{\eta }} \right)^N. \label{outage0}
\end{equation}
The right-hand-side (RHS) of (\ref{outage0}) is the outage
probability for MRC-TB when $K=1$, while the left-hand-side (LHS)
expression is the outage probability for space-time coding for any
$K$.
\end{lemma}
\begin{proof}
The proof can be completed by using the inequality $\rho
_d^{\left(N\right)} \ge \rho _d^{\left\{m_k\right\}} \ge
\rho_{std}$, and the following approximation \cite{fund}:
\begin{equation}
P\left( {\sum\limits_{k = 1}^K {\sum\limits_{i = 1}^{m_i } {\left|
{g_{i,k} } \right|^2 }  \le \varepsilon } } \right) \approx
\frac{1}{{N!}} \varepsilon ^N. \label{approxchi}
\end{equation}
Further details are omitted due to space limitations.
\end{proof}
\emph{Lemma 2} indicates that the full diversity of $N$ can be
achieved regardless of the number of relays $K$, provided that the
signals are correctly decoded at the relays. However, the diversity
of the network might decrease if decoding outages occur at the
relays. To avoid this event, we need to select only the relays that
can decode the signal correctly. In fact, we can extend the antenna
selection protocol proposed by \cite{23}, which exploits further the
selection diversity of the source to relay channels, to the
multi-antenna multi-relay scenario discussed in the \emph{letter},
as follows.
\newtheorem{protocol}{Protocol}
\begin{protocol}
\textbf{(Selection Decoding)} In order to decode and forward the
messages, select ${\tilde K}$ relays with a total number of ${\tilde
N}$ antennas, denoted as a set ${\Re \left( {\tilde N,\tilde K}
\right)}$, that can successfully decode the source message at a
transmission rate $R$.
\end{protocol}
We can obtain the outage probability for selection decoding in the
following theorem:
\begin{theorem}
For large $\eta$, the outage probability for the selection decoding
scheme for \emph{any} $K$ and $\left\{m_k\right\}$ is bounded
approximately by:
\begin{eqnarray}
\left( {\frac{{2^{2R}  - 1}}{\eta }} \right)^N \sum\limits_{\Re
\left( {\tilde N,\tilde K} \right)} && {\left( {\frac{{\tilde
N^{\tilde N} }}{{\tilde N!}}\prod\limits_{r \notin \Re \left(
{\tilde N,\tilde K} \right)} {\frac{1}{{m_r !}}} } \right)}  \ge
P_{out}^{\left\{m_k\right\} } \ge \nonumber \\  \left(
{\frac{{2^{2R} - 1}}{\eta }} \right)^N && \sum\limits_{\Re \left(
{\tilde N,\tilde K} \right)} {\left( {\frac{1}{{\tilde
N!}}\prod\limits_{r \notin \Re \left( {\tilde N,\tilde K} \right)}
{\frac{1}{{m_r !}}} } \right)}, \label{selcod1}
\end{eqnarray}
where the RHS is achieved for MRC-TB when $\tilde N$ antennas are
co-located (i.e., $\tilde K$ relays cooperate like one
relay\footnote{This implies that the selected relays can always
jointly decode and jointly transmit as if they were one relay.
Therefore it is an idealistic case and thus can be only considered
as a performance upper bound}), the LHS represents the outage
probability for the space-time coding scheme.
\end{theorem}

\begin{proof}
See Appendix.
\end{proof}

\begin{figure}[t!]
\centering
\includegraphics[width=3.5in]{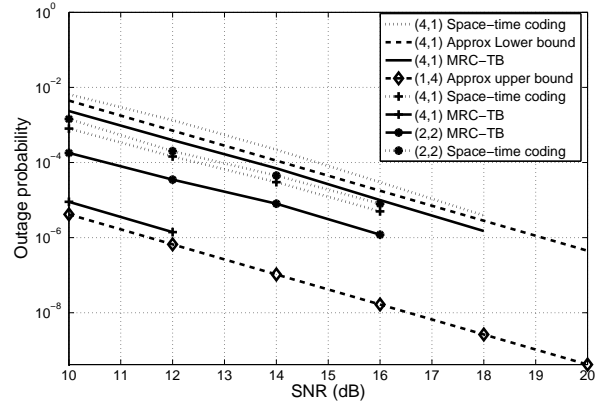}
\caption{Outage probability for different pairs of $(K,M)$, where
$K$ is the number of relays and $M$ is the number of antennas at
each relay. Dashed lines are approximations for high SNR using
(\ref{selcod1}). Dotted curves are simulations for space-time
coding. Solid curves are simulations for MRC-TB.} \label{dmt}
\end{figure}

It can be seen from \emph{Theorem 1} that for selection decoding
full diversity can always be achieved regardless of the choices of
$K$ and $\left\{m_k\right\}$, and the performance is lower bounded
by that of the space-time coding scheme. However, it can also be
seen that different choices of $K$ and $\left\{m_k\right\}$ might
result in different performance, due to different power gains.
Comparing the RHS and LHS of (\ref{selcod1}), we can see a factor of
$\tilde N ^{\tilde N}$, where $\tilde N$ can be any value from $0$
to $N$. This implies that the performance gap between MRC-TB and
space-time coding can be extremely large when $N$ is large. Note
that in practice a large $N$ (i.e, the number of transmit antennas)
might not be realistic for point-to-point MISO link, and therefore
the performance advantage for TB is always limited. However, in a
large ad-hoc or sensor network, it is quite possible to have large
values of $N$, and thus the benefits of distributed MRC-TB can be
significant. Furthermore, the benefits of deploying multiple
antennas at the relays (i.e., applying MRC at the relays) is small
when distributed space-time coding is used, as the performance is
mainly constrained by the limited power gain of using space-time
coding for the relay to destination link. TB in this scenario can
offer significant performance advantages. Fig. 2 shows a simulation
example for $N=4$. It can be observed that the performance gap
between MRC-TB and space-time coding becomes largest when all the
antennas are deployed at a single relay (a $6dB$ difference in this
example).

We further note that, in practice, in order to achieve full
diversity gain, the relay selection protocol is easier to implement
for distributed MRC-TB than for space-time coding. The reason is
that for space-time coding, the codes (e.g., block length or code
pattern) must be changed whenever the number of selected relays are
changed, in order to obtain the full diversity. This will involves
much more channel feedback and signaling overhead.


%

%

%

\section{Conclusions}

The performance of the distributed MRC-TB scheme has been studied in
a multi-antenna multi-relay environment. We have seen that this
technique achieves full diversity regardless of the number of relays
and antennas at each relay, and offers a significant power gain over
space-time coding.

Note that two important issues about the MRC-Beamforming approach
are synchronization and frequency offset among all the relays
(\cite{cps,dstc,acd}). The impact of these two issues on the relay
network is an interesting topic for future research.


%
%

\appendix[Proof of Theorem 3]

Since ${\Re \left( {\tilde N,\tilde K} \right)}$ is a random set, we
use the law of total probability and write
\begin{equation}
P_{out}  = \sum\limits_{\Re \left( {\tilde N,\tilde K} \right)}
{P\left[ {\Re \left( {\tilde N,\tilde K} \right)} \right]}
P_{out}^{m_k |\Re \left( {\tilde N,\tilde K} \right)},
\label{outsel}
\end{equation}
where $P_{out}^{m_k |\Re \left( {\tilde N,\tilde K} \right)}$
denotes the outage probability conditioned on the event that ${\Re
\left( {\tilde N,\tilde K} \right)}$ is chosen, and can be bounded
by (\ref{outage0}) by replacing $N$ with $\tilde N$. The probability
that any relay is chosen can be expressed as
\begin{eqnarray}
P\left[ {r \in \Re \left( {\tilde N,\tilde K} \right)} \right] &=&
P\left[ {\sum\limits_{i = 1}^{m_k } {\left| {h_{i,k} } \right|^2 }
\ge \frac{{2^{2R}  - 1}}{\eta }} \right] \nonumber \\  &=& 1 -
P\left[ {\sum\limits_{i = 1}^{m_k } {\left| {h_{i,k} } \right|^2 }
\le \frac{{2^{2R}  - 1}}{\eta }} \right].
\end{eqnarray}
Therefore a set ${\Re \left( {\tilde N,\tilde K} \right)}$ exists
with a probability that can be written as
\begin{eqnarray}
P\left[ {\Re \left( {\tilde N,\tilde K} \right)} \right] &=&
\prod\limits_{r \in \Re \left( {\tilde N,\tilde K} \right)} {\left(
{1 - P\left[ {\sum\limits_{i = 1}^{m_k } {\left| {h_{i,k} }
\right|^2 }  \le \frac{{2^{2R}  - 1}}{\eta }} \right]} \right)}
\nonumber \\ & & {} \times \prod\limits_{r \notin \Re \left( {\tilde
N, \tilde K} \right)} {P\left[ {\sum\limits_{i = 1}^{m_k } {\left|
{h_{i,k} } \right|^2 } \le \frac{{2^{2R}  - 1}}{\eta }} \right]}.
\end{eqnarray}
Based on (\ref{approxchi}), at high SNR, $P\left[ {\Re \left(
{\tilde N,\tilde k} \right)} \right]$ can be approximated as
\begin{equation}
P\left[ {\Re \left( {\tilde N,\tilde K} \right)} \right] \approx
\left( {\frac{{2^{2R}  - 1}}{\eta }} \right)^{N - \tilde N}
\prod\limits_{r \notin \Re \left( {\tilde N,\tilde K} \right)}
{\frac{1}{{m_k !}}}.\label{pset}
\end{equation}
Putting (\ref{pset}) and (\ref{outage0}) into (\ref{outsel}), we
obtain the bound in (\ref{selcod1}) and thus the proof is complete.

\end{document}